\documentclass[prl,twocolumn,showpacs,superscriptaddress]{revtex4-1}

\usepackage{amsmath,amsthm,amsfonts,amssymb}
\usepackage{epsf}
\usepackage{graphicx}
\newcommand{\rv}{\mathbf{r}}
\newcommand{\Rv}{\mathbf{R}}
\newcommand{\eps}{\epsilon}
\newcommand{\mone}{{-1}} 
\newcommand{\brho}{\mbox{\boldmath$\rho$}}
\newcommand{\kv}{\mathbf{k}}
\newcommand{\qv}{\mathbf{q}}
\newcommand{\G}{\mathbf{G}}

\begin{document}
\title{Dielectric screening in extended systems \\
       using the self-consistent Sternheimer equation and localized basis sets}

\author{Hannes H\"ubener}
\affiliation{Department of Materials, University of Oxford, Oxford OX1 3PH, United Kingdom}

\author{Miguel A. \surname{P\'erez-Osorio}} 
\affiliation{Department of Materials, University of Oxford, Oxford OX1 3PH, United Kingdom}
\affiliation{Centre d'Investigaci\'o en Nanoci\`encia i Nanotecnologia-CIN2
(CSIC-ICN), Campus UAB, Bellaterra, Spain}

\author{Pablo Ordej\'on}
\affiliation{Centre d'Investigaci\'o en Nanoci\`encia i Nanotecnologia-CIN2
(CSIC-ICN), Campus UAB, Bellaterra, Spain}

\author{Feliciano Giustino}
\affiliation{Department of Materials, University of Oxford, Oxford OX1 3PH, United Kingdom}

\begin{abstract}
We develop a first-principles computational method for investigating the dielectric 
screening in extended systems using the self-consistent Sternheimer equation and
localized non-orthogonal basis sets. Our approach does not require the explicit
calculation of unoccupied electronic states, only uses two-center integrals, and 
has a theoretical scaling of order $O(N^3)$. We demonstrate this method by comparing
our calculations for silicon, germanium, diamond, and LiCl
with reference planewaves calculations. We show that accuracy comparable to
planewaves calculations can be achieved via a systematic optimization of the
basis set.
\end{abstract}

\date{\today{ }}

\pacs{71.15.-m   
      71.15.Ap   
      71.15.Dx}  

\maketitle

Quasiparticle calculations based on the $GW$
method \cite{hedin65,hedin1969,hybertsen86,aryasetiawan98,aulbur2000,onida02}
and calculations of optical
spectra using the Bethe-Salpeter equation \cite{onida95,rohlfing98,onida02}
have emerged as state-of-the-art first-principles computational methods for 
the study of electronic and optical excitations both in extended solids and 
in nanoscale systems. While in many cases these methods can reproduce and even 
predict experimental photoemission and optical data with remarkable accuracy, 
the associated computational complexity limits their application to relatively 
small systems, typically in the range of ten to a hundred atoms.
The main bottleneck in these calculations is the evaluation of the irreducible 
polarization propagator in the random-phase approximation (RPA), which is used for
constructing the screened Coulomb interaction of the $GW$ self-energy,
and the electron-hole interaction kernel in the Bethe-Salpeter equation.
The calculation of this propagator is performed using an expansion over unoccupied
Kohn-Sham states 
\cite{hybertsen86,rohlfing93,rohlfing95,aryasetiawan98,aulbur2000,onida02,
blase11}, 
and has a theoretical scaling of 
$O(N^4)$, $N$ being the number of atoms in the system \cite{giustino10}
\footnote{By using planewaves basis sets and fast Fourier transform techniques
a scaling $O(N^3{\rm log}N)$ can be obtained.}.
In addition the polarization propagator is commonly calculated by expanding
the Kohn-Sham states using planewaves basis sets \cite{ihm79}.
While planewaves have several obvious advantages, large systems can become 
intractable due to the very large basis size.
These observations clearly point to the need for new methods to study 
the dielectric screening in complex systems which (i) do not require the
explicit calculation of unoccupied electronic states, (ii) use small
basis sets, and (iii) exhibit a more favorable scaling with system size.

Several methods have been proposed in order to avoid the explicit calculation
of unoccupied electronic states in the study of electronic 
excitations \cite{levine89,levine91,reining97,andrade07,wilson08,bruneval08,
wilson09,rocca10,giustino10,umari10,berger10}. Some of these methods rely on
the Sternheimer
equation \cite{levine89,levine91,reining97,andrade07,giustino10}, 
which is routinely used in density-functional perturbation
theory \cite{baroni01},
and closely related to the coupled-perturbed Hartree-Fock method
found in the quantum chemistry literature \cite{mcweeny62,gerratt68,Liang05}.
In particular, in Ref.~\cite{giustino10} the Sternheimer method is used 
self-consistently in order to calculate the full frequency-dependent inverse 
dielectric function in extended systems, and an empirical pseudopotential 
implementation is described as a proof-of-concept.

In order to reduce the computational complexity of Sternheimer-type approaches
it appears advantageous to adapt these methods to the case of localized basis sets. 
Indeed, electronic structure codes based on localized representations are
successfully 
employed to study very large systems, up to several thousands of 
atoms \cite{soler02,skylaris05,bowler10}. 
The key advantage of these basis sets is that they exploit the electron
localization in the insulating state \cite{kohn96} in order to obtain
a sparse representation of the ground-state density matrix.
However, since these basis sets are optimized for providing an accurate 
description of the occupied Kohn-Sham manifold, 
it is not clear {\it a priori} what their performance would be in the case
of excited state calculations.

In this work we demonstrate a first-principles pseudopotential scheme 
for calculating the inverse dielectric matrix in extended systems using 
non-orthogonal localized basis sets. Our scheme includes both local-field effects 
and the frequency-dependence of the inverse dielectric matrix, and does not 
require the explicit calculation of unoccupied Kohn-Sham states. This is
achieved by adapting the self-consistent Sternheimer 
method of Ref.~\cite{giustino10} to the case of a basis set of localized
pseudo-atomic orbitals as implemented in the {\tt SIESTA}
code \cite{portal96,artacho99,soler02}.
Here we illustrate the formalism and report calculations of the inverse
dielectric
matrix for prototypical semiconductors and insulators. We perform a systematic
comparison with reference planewaves calculations, and we investigate the
convergence of our results with the size of the local orbital basis set.
A detailed report with extensive benchmarks and implementation details is
provided elsewhere \cite{huebener12}.

We calculate the inverse dielectric matrix in the random-phase approximation
following Ref.~\cite{giustino10}. The frequency-dependent variation 
of the density matrix $\Delta n(\rv,\rv',\omega)$ corresponding to 
the non-local bare Coulomb potential $v(\rv,\rv')$ is given by:
  \begin{equation}\label{eq:den}
  \Delta n(\rv,\rv',\omega) = 2\!\sum_{v,\sigma=\pm}\psi^*_v(\rv')
  \Delta\psi_v(\rv,\rv',\sigma\omega),
  \end{equation}
where $\psi_v$ is an occupied Kohn-Sham state of energy $\eps_v$ and
spin-unpolarized systems are considered for simplicity.
The variations $\Delta\psi_v(\rv,\rv',\pm\omega)$ of the single-particle
states are solutions of the Sternheimer equations:
  \begin{equation}\label{eq:stern}
  (\hat{H}-\eps_v \pm \omega)\Delta \psi_v(\rv,\rv',\pm\omega) =
  -(1-\hat{P}_{v})W(\rv,\rv',\omega)\psi_v(\rv'),
  \end{equation}
where Kohn-Sham Hamiltonian $\hat{H}$ and the projector $\hat{P}_v$ on the
occupied manifold act on the variable $\rv'$, and
$W(\rv,\rv',\omega)$ is the RPA frequency-dependent screened Coulomb interaction.
In order to obtain the screened Coulomb interaction we use the
change $\Delta V_H$ of the Hartree potential resulting from the
variation of the density matrix:
  \begin{eqnarray}
  \label{eq:hart} 
  \Delta V_H(\rv,\rv',\omega) &=& \int d\rv''\Delta
  n(\rv,\rv'',\omega)v(\rv'',\rv'), \\
  \label{eq:pot}
  W(\rv,\rv',\omega) &=& v(\rv,\rv') + \Delta V_H(\rv,\rv',\omega).
  \end{eqnarray}
Equations (\ref{eq:den})-(\ref{eq:pot}) need to be solved self-consistently. 
In the following we use this formalism in order to directly calculate the inverse
dielectric matrix $\eps^\mone(\rv,\rv',\omega)$. This is done by
replacing $W$ in Eq.~(\ref{eq:stern}) by $\eps^\mone$, and by using
the following relation instead of Eq.~(\ref{eq:pot}):
  \begin{equation}
  \eps^\mone(\rv,\rv',\omega) = \delta(\rv,\rv')+ \Delta V_H(\rv,\rv',\omega) .
  \label{eq:eps}
  \end{equation}
In this case the self-consistent calculation is started
by initializing the inverse dielectric matrix to the Dirac delta $\delta(\rv,\rv')$.
A derivation of the connection with the sum-over-states approach
and details on this formalism can be found in Refs.~\cite{giustino10,huebener12}
respectively.

We now move to a localized basis representation. We first make the choice
of treating the variables $\rv$ and $\omega$ in the Sternheimer equation 
Eq.~(\ref{eq:stern}) as parameters. The variable $\rv$ will be represented
on a real-space grid. Then we expand the quantities dependent on $\rv'$ 
in the basis of local orbitals $\phi_i(\rv')$:
  \begin{eqnarray}\label{eq:expVal}
  \psi_v(\rv') &=& \sum_i c_{v i} \phi_i(\rv')\\
  \Delta\psi^\pm_{v[\rv,\omega]}(\rv') &=& \sum_i \Delta c^\pm_{v i[\rv,\omega]}
  \phi_i(\rv') \label{def:dc}.
  \end{eqnarray}
By replacing Eqs.~(\ref{eq:expVal}),(\ref{def:dc}) in Eq.~(\ref{eq:stern})
and projecting both sides onto $\phi_j(\rv')$ we obtain the Sternheimer equation
in matrix representation:
  \begin{equation}\label{eq:sternMatrix}
  \left[\mathbf{H} - (\epsilon_v\pm \omega)\mathbf{S}\right]\Delta
  \mathbf{c}^\pm_{v[\rv,\omega]} = -\left[\mathbf{1}
  -\mathbf{S}\brho^T \right]\Delta \mathbf{V}_{[\rv,\omega]} \mathbf{c}_v.
  \end{equation}
Here $\mathbf{H}$, $\mathbf{S}$ and $\brho$ are the Kohn-Sham Hamiltonian,
the overlap, and the density matrices, respectively. The matrix elements
of $\Delta \mathbf{V}_{[\mathbf{r},\omega]}$ in Eq.~(\ref{eq:sternMatrix}) 
are defined as:
  \begin{equation} \label{eq:dVint}
  \Delta V_{ij[\rv,\omega]} = \int d\rv' \phi^*_i(\rv')  \Delta
  V_{[\rv,\omega]}(\rv') \phi_j(\rv') .
  \end{equation} 
The solution of Eq.~(\ref{eq:sternMatrix}) yields the coefficients
$\Delta \mathbf{c}^\pm_{v[\rv,\omega]}$ which are used to construct the
variation of the density matrix from Eq.~(\ref{eq:den}):
  \begin{equation}
  \Delta n_{ij[\rv,\omega]} = 2\sum_{v,\sigma=\pm} c^*_{v i}\Delta
  c^\sigma_{v j[\rv,\omega]}.
  \end{equation}
At this point it is possible to explicitly calculate 
$\Delta n(\rv,\rv',\omega)$ on a real-space grid, and then
perform the integral in Eq.~(\ref{eq:hart}) on the same grid.
Equation~(\ref{eq:pot}) is also evaluated on the real space grid,
and the resulting potential is projected on the local orbital basis
using Eq.~(\ref{eq:dVint}).
This procedure is repeated until self-consistency is achieved, and is
carried out independently for each value of the parameters $\rv$ and $\omega$.

For the purpose of comparison with standard planewaves calculations
it is convenient to specialize this formalism to the case of 
crystalline solids. We proceed as follows: we incorporate
the Bloch phase factors in the basis orbitals $\phi_{i\kv}(\rv')$
  \begin{equation}\label{def:kbasis}
  \phi_{i\kv}(\rv') = \sum_{\Rv}e^{-i\kv\cdot(\rv-\Rv)}\phi_{i}(\rv'-\Rv),
  \end{equation}
where $\kv$ is a wavevector in the Brillouin zone, and $\Rv$ is a lattice vector.
The orbitals $\phi_{i}$ in Eq.~(\ref{def:kbasis}) are the same as 
in Eq.~(\ref{eq:expVal}), except that now the index $i$ runs over the
orbitals spanning one unit cell only.
Equation~(\ref{def:kbasis}) defines a basis of periodic functions and
is used to expand the cell-periodic part $u_{v\kv}$ of the Bloch states
$\psi_{v\kv}(\rv') =  e^{i\kv\cdot\rv'}u_{v\kv}(\rv')$. By using a similar
expansion for the variations of the wavefunctions in Eq.~(\ref{def:dc}), 
and by projecting both sides of Eq.~(\ref{eq:stern}) on the basis orbitals, 
we obtain the Sternheimer equation for periodic systems:
  \begin{equation}\label{eq:ksternMatrix}
  \begin{split}
  &\left[\mathbf{H}_{\kv+\qv} - (\epsilon_{v\kv}\pm
  \omega)\mathbf{S}_{\kv+\qv}\right]\Delta \mathbf{c}^\pm_{v\kv
  [\qv,\rv,\omega]} = \\
  & \qquad\qquad\; -\left[\mathbf{1} -
  \mathbf{S}_{\kv+\qv}\mathbf{\brho}^T_{\kv+\qv}\right]\Delta
  \mathbf{V}_{\kv[\qv,\rv,\omega]} \mathbf{c}_{v\kv},
  \end{split}
  \end{equation}
which is analogous to Eq.~(\ref{eq:sternMatrix}). In this case the Hamiltonian,
overlap, and density matrices, as well as the coefficient vectors, are all resolved 
in momentum space. A detailed derivation of these equations is provided
in Ref.~\cite{huebener12}.

In the present formalism the dependence on the real-space variable $\rv'$
is represented on the local orbital basis, while the dependence on $\rv$
is represented on a real-space grid. This choice carries the following
advantages: (i)~we do not use a product-basis expansion for non-local
quantities, therefore we avoid issues related to the representability of
the screened Coulomb interaction in localized basis sets. (ii)~We only have
two-center integrals in our formulation, therefore we do not need the
three and four-center integrals arising in product basis
expansions \cite{rohlfing95,blase04,koval10,foerster11,blase11}. (iii)~The
self-consistent 
calculation of the inverse dielectric matrix avoids from the outset
the issues associated with the inversion of response functions represented
in local orbital basis sets \cite{gidopoulos11}. (iv)~Since
Eq.~(\ref{eq:sternMatrix})
can be solved using sparse linear algebra and needs to be performed
for every occupied state and every $\rv$ on the real-space grid,
this method has a theoretical scaling of~$O(N^3)$.

The key approximation in our formulation is the expansion of the variation
$\Delta\psi^\pm_{v[\rv,\omega]}$ of the single-particle states in the basis 
of local orbital [cf.~Eq.(\ref{def:dc})]. In fact local orbital basis sets
are typically optimized to accurately describe the occupied states manifold,
while the variation of the density matrix arises from the components of
$\Delta\psi^\pm_{v[\rv,\omega]}$ in the manifold of unoccupied
states \cite{baroni01}.
This clearly points to the need of carefully optimizing
the local orbital basis sets for calculations of the dielectric screening.
A systematic assessment of the performance of multiple-$\zeta$ polarized
pseudo-atomic orbital basis sets is provided in Ref.~\cite{huebener12}. 

Our implementation is based on the {\tt SIESTA} code, and uses a basis of strictly
localized numerical pseudo-atomic orbitals \cite{portal96,artacho99,soler02}.
In the following we discuss benchmark results for silicon, diamond, germanium and
LiCl. Calculations are performed within the local-density approximation (LDA)
to 
density-functional theory \cite{ceperley80,perdew81}, and with
norm-conserving pseudopotentials \cite{troullier91}.
The inverse dielectric matrices are calculated by sampling the Brillouin
zone using a shifted 10$\times$10$\times$10 grid for diamond, silicon
and LiCl, and a shifted 12$\times$12$\times$12 grid for germanium.
We use the lattice parameters 5.43 \AA, 3.56 \AA, 5.65 \AA, and 5.13 \AA\
for silicon, diamond, germanium, and LiCl, respectively \cite{hybertsen87}. 
Using a standard triple-$\zeta$ polarized (TZP) basis we obtain
direct band gaps of 2.55 eV (Si), 5.60 eV (diamond), 0.04 eV (Ge), 5.99 eV
(LiCl),
in line with standard LDA results.
In order to generate basis sets for silicon with a large number of $\zeta$'s we
use
an energy-shift parameter \cite{artacho99} of 10 meV.
This shift leads to localization radius of 9.3 \AA\ for silicon which is
slightly larger
than those adopted in standard ground-state calculations using {\tt SIESTA}.
We compare our results with planewaves calculations performed using
{\tt ABINIT} \cite{abinit} and {\tt YAMBO} \cite{yambo} software packages,
with same the pseudopotentials and Brillouin-zone sampling for
consistency \footnote{ We note that there is a small difference between {\tt
SIESTA} and {\tt ABINIT} in how the local part of the pseudopotential is 
constructed.}.
Planewaves calculations are carried out using kinetic energy cutoffs 
of 20 Ry for Si and Ge, and of 60 Ry for diamond and LiCl. The RPA dielectric
matrices are calculated using 92 conduction bands and
kinetic energy cutoffs of 6.9 Ry for silicon, germanium and LiCl, and of 12 Ry 
for diamond.

\begin{figure}
  \begin{center}
  \resizebox{0.9\columnwidth}{!}{\includegraphics{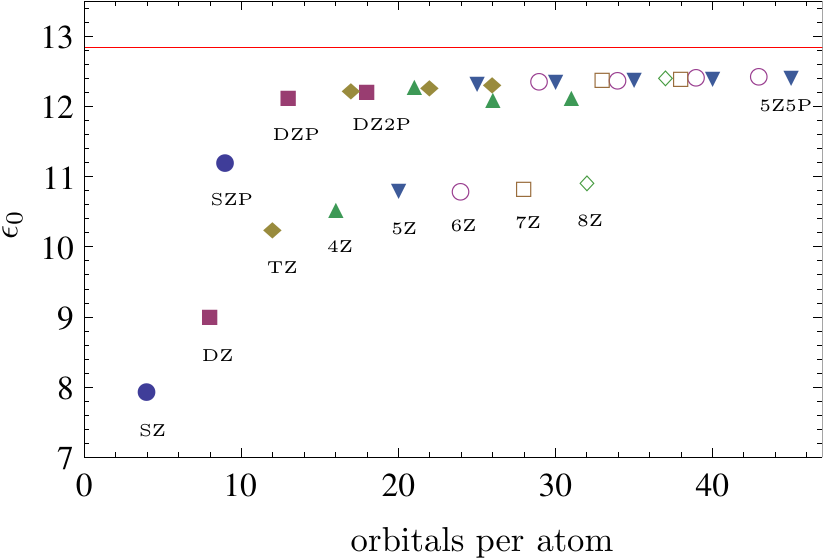}}
  \end{center}
  \caption{
  Calculated macroscopic dielectric constant of silicon 
  $\eps_0 = 1/\eps^{-1}_{{\bf 0}{\bf 0}}({\bf q}\rightarrow 0,\omega=0)$
  as a function of basis size, given in terms of orbitals per atom
  [we use ${\bf q} = 2\pi/a (0.01,0,0)$, $a$ being the Si lattice constant].
  The number of $\zeta$ functions included in each basis set is indicated 
  by the labels SZ, DZ, TZ, etc. The number of polarization orbitals 
  included in each basis set increases towards the right-hand side, as indicated
  for the case of the DZ basis. The reference planewaves calculation is indicated
  by the horizontal line.
  \label{fig1}}
  \end{figure}

Figure~\ref{fig1} shows the calculated static macroscopic dielectric constant of 
silicon as a function of basis set size. The size of the basis is increased by including
additional $\zeta$'s using the split-norm procedure, as well
as polarization orbitals \cite{artacho99}.
We find that the single-$\zeta$ polarized (SZP) basis already gives results
which are within 10\% of the reference planewaves calculation. This suggests
that a suitably optimized basis with a few orbitals should be able to yield
converged results. We also observe a monotonic convergence with the number
of basis orbitals, and a consistently superior performance of the basis sets
including polarization orbitals (Fig.~\ref{fig1}).

\begin{figure}
  \centering
  \resizebox{0.9\columnwidth}{!}{\includegraphics{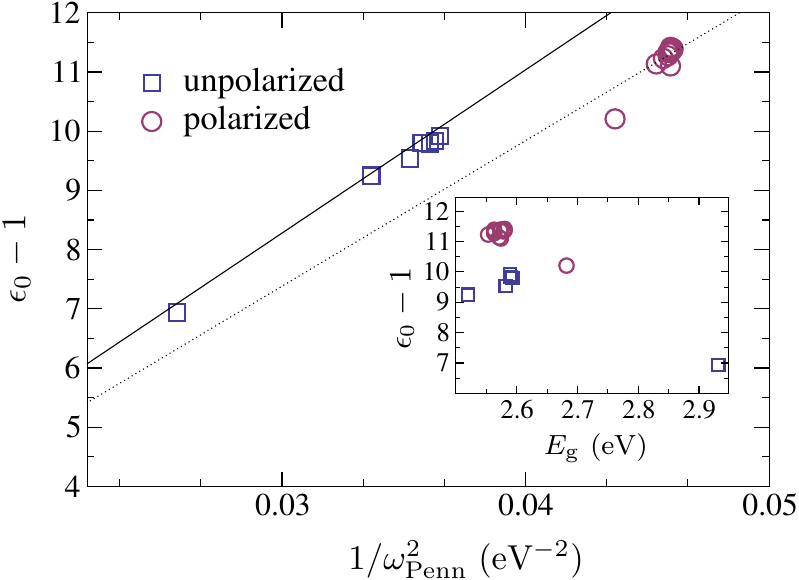}}
  \caption{
  Macroscopic dielectric constant of silicon vs. Penn gap, plotted as
  $\eps_0-1$ vs. $1/E_{\rm P}^2$, with unpolarized (squares) and polarized
(circles) 
  basis sets. The slope of the solid line corresponds to the square of the
  plasma frequency of silicon, the dotted line is a guide to the eye. 
  Inset: macroscopic dielectric constant vs. the direct band gap
  $E_{\rm g}$.}
  \label{fig2}
  \end{figure}

We can rationalize the convergence of the dielectric constant with basis size
shown in Fig.~\ref{fig1} by considering the Penn model for the dielectric
constants of semiconductors \cite{penn62}.
In this simple model the dielectric constant is related to the Penn gap 
$E_{\rm P}$ and to the plasma frequency $\omega_{\rm p}$ 
by $\epsilon_0 = 1 + (\hbar\omega_{\rm p}/E_{\rm P})^2$.
In our calculations we extract the Penn gap for each basis set considered 
in Fig.~\ref{fig1} using the main peak in the joint density of states
(JDOS) \cite{cardona} \footnote{
   For a reference basis set we choose the Penn gap as the energy of the
   main peak of the JDOS.
   The cumulative integral of the JDOS from the absorption threshold to this 
   energy is then used to define unambiguously the Penn gap for
   the other basis sets.}. 
Figure~\ref{fig2} shows that $\eps_0-1$ is approximately a linear function
of $1/E_{\rm P}^2$. In particular the values calculated using the unpolarized
basis sets nicely fall on the line obtained using the plasma frequency of
silicon $\hbar\omega_{\rm p}$ =16.6 eV.
For comparison, the inset of Fig.~\ref{fig2} shows that the calculated
dielectric constants do not correlate with the direct band gap of silicon.
These observations lead us to conclude that the Penn gap, and more generally 
the low-energy structure of the JDOS, is a good indicator of the quality
of the basis set for calculating dielectric constants.
This is helpful for a quick assessment of the quality of a basis set without
explicitly calculating the screening.

Table~\ref{tab:G} shows the first few elements of the inverse dielectric matrix
$\eps^{-1}_{{\bf G}{\bf G}'}({\bf q}\rightarrow 0,\omega=0)$ for several local
orbital 
basis sets. We observe that the calculated values for the first few
shells compare well with the reference planewaves calculation, similarly to
the head of the dielectric matrix. For higher shells, however, the relative agreement 
of the basis sets worsens slightly. This may indicate a systematic deficiency 
in the local orbital basis in case of the higher Fourier components of 
dielectric matrix, and clearly deserves further investigation.
  \begin{table}
  \begin{center}
  \begin{tabular}{cc|rrr|r}
  \hline
  \hline
  $\G$   & $\G'$        & SZ          & TZP         & CZ4P              & PW \\\hline
  (0,0,0) & (0,0,0) \quad&\quad  0.125 & \quad 0.081 & \quad 0.080  \quad&\quad 0.077   \\
  (1,1,1) & (1,1,1) \quad&\quad  0.822 &\quad  0.617 &\quad  0.600  \quad&\quad  0.637  \\
  (1,1,1) & (2,0,0) \quad&\quad -0.037 &\quad -0.040 &\quad -0.036  \quad&\quad -0.038  \\
  (2,0,0) & (2,0,0) \quad&\quad  0.839 &\quad  0.686 &\quad  0.664  \quad&\quad  0.766  \\
  (2,2,2) & (2,2,2) \quad&\quad  0.984 &\quad  0.952 &\quad  0.937  \quad&\quad  0.984  \\
  (2,2,2) & (1,1,1) \quad&\quad -0.009 &\quad -0.030 &\quad -0.034  \quad&\quad -0.020  \\
  (2,2,2) & (2,0,0) \quad&\quad -0.003 &\quad  0.003 &\quad  0.005  \quad&\quad  0.002  \\
  \hline
  \hline
  \end{tabular}
  \end{center}
  \caption{
  Fourier components of the symmetrized inverse dielectric matrix
  $\eps^{-1}_{{\bf G}{\bf G}'}({\bf q}\rightarrow 0,\omega=0)$ of silicon,
calculated
  using the SZ basis (4 orbitals per atom), TZP (17 orbitals per
  atom), the CZ4P basis (40 orbitals per atom), and a planewaves (PW) basis set.
  The reciprocal lattice vectors are given in units of $2\pi/a$, $a$ being
  the Si lattice parameter.
  \label{tab:G}}
  \end{table}

\begin{figure}
  \resizebox{0.9\columnwidth}{!}{\includegraphics{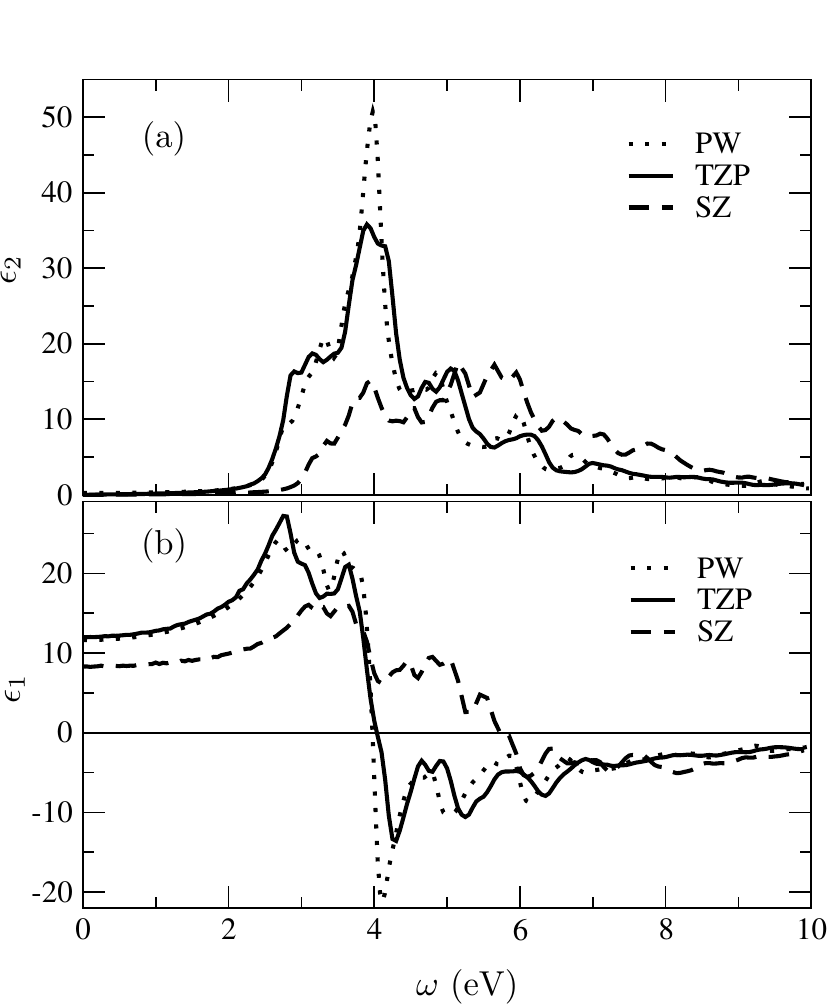}}
  \caption{
  Macroscopic dielectric function of silicon 
  $\epsilon(\omega)=\eps^{-1}_{{\bf 0}{\bf 0}}({\bf q}\rightarrow 0,\omega)$ as
  a function
  of frequency: (a) imaginary part $\eps_2(\omega)$, and (b) real part $\eps_1(\omega)$.
  We compare our calculations performed using the SZ basis (dashed lines) and the
  TZP basis (solid lines) with reference planewaves calculations (dotted lines).
  We use an energy broadening of 0.1 eV.}
  \label{fig3}
  \end{figure}

Figure~\ref{fig3} shows the frequency-dependent dielectric function of
silicon calculated using our method, together with a reference planewaves
calculation. We observe that, as expected, the performance of the minimal
single-$\zeta$ basis is poor. In fact, spectral 
weight is incorrectly transferred from the main absorption peak to higher energies.
On the contrary, the TZP basis (17 orbitals per atom) performs quite well, 
with all the main features of the planewaves spectra correctly reproduced. 
We still observe however some small transfer of spectral weight and a slight 
blueshift of the high-energy peaks. 

In order to demonstrate the generality of of our approach we present in
Tab.~\ref{tab1} our calculated dielectric constants for germanium, 
diamond, and LiCl. The values shown in Tab.~\ref{tab1} are obtained by
using standard {\tt SIESTA} settings for the basis \footnote{In this table we
use
a split-norm of 0.15 and an energy shift of 200 meV for
silicon, diamond and germanium, and end nergy shift of 70 meV for LiCl.
This choice leads to small localization radii, for instance we obtain
a radius of 6.4 \AA\ in the case of Si.}.
Table~\ref{tab1} provides further support to our previous finding by showing
that the TZP basis provides results which lie within 1-5\% of the reference
planewaves calculations. We expect further improvement upon designing
basis sets specifically optimized for the Sternheimer scheme proposed in this
work. In particular the use of numerical diffuse
orbitals \cite{cofini07,ordejon09} deserves a systematic assessment.

  \begin{table}
  \begin{center}
  \begin{tabular}{l|rrrr}
  \hline\hline
          &    SZ  &   TZP  &    PW & Expt. \\ 
  \hline
  Si      &  8.41  & 12.25  & 12.84 & 11.7$^{\rm a}$\\
  Ge      & 15.38  & 18.15  & 17.92 & 15.8$^{\rm a}$ \\
  Diamond &  3.94  &  5.43  &  5.47 & 5.5$^{\rm a}$ \\
  LiCl    &  1.69  &  2.68  &  2.82 & 2.8$^{\rm b}$ \\
  \hline
  \hline
  \end{tabular} \\
  \hspace{-3.5cm}
  $^{\rm (a)}$ Ref.~\cite{kittel}. \\
  \hspace{-3.5cm}
  $^{\rm (b)}$ Ref.~\cite{ashcroft}. 
  \end{center}
  \caption{
  Dielectric constants of Si, Ge, diamond, and LiCl calculated
using our
  Sternheimer approach, and compared to reference planewaves calculations
  and experimental data. We report both our results obtained using the 
  minimal SZ basis and the TZP basis, as generated using
  the standard settings of {\tt SIESTA}.
  \label{tab1}}
  \end{table}

In conclusion, we have introduced and demonstrated a method for calculating
the inverse dielectric matrix of extended systems combining localized 
non-orthogonal basis sets with the self-consistent Sternheimer equation.
Our method does not require the calculation of unoccupied electronic states, 
does not require the explicit inversion of the dielectric matrix, 
uses only two-center integrals, and has a theoretical scaling with system size 
of~$O(N^3)$. Our implementation based on the pseudo-atomic orbitals of the {\tt
SIESTA} 
code shows that results with accuracy comparable to planewaves calculations
can be obtained upon optimization of the localized basis set.
Our formulation is completely general and there should be no difficulties
in adapting our method to other local basis implementations.
We believe this work represents a stepping stone towards quasiparticle 
$GW$ calculations for large and complex systems using localized basis sets.

The authors would like to thank E. Artacho for fruitful discussions. This work
is funded by the European Research Council under the European
Community's Seventh Framework Programme, Grant No. 239578, and Spanish MICINN
Grant FIS2009-12721-C04-01.

%

\end{document}